\documentclass[twocolumn,amsmath,showpacs,amsfonts,aps,prc,floatfix]{revtex4}
\usepackage{graphicx}
\usepackage{bm}
\newcommand{\la}{\langle}
\newcommand{\ra}{\rangle}
\begin{document}
%%\draft
\title{Large elliptic flow in low multiplicity pp collisions at LHC energy $\sqrt{s}$=14 TeV}
 
\author{A. K. Chaudhuri}
\email[E-mail:]{akc@veccal.ernet.in}
\affiliation{Variable Energy Cyclotron Centre, 1/AF, Bidhan Nagar, 
Kolkata 700~064, India}

\begin{abstract}

We explore the possibility of observing elliptic flow in low multiplicity events in central pp collisions at LHC energy, $\sqrt{s}$=14 TeV. It is
assumed that the initial interactions produces a number of hot spots. Hydrodynamical evolution of two or more hot spots can   generate sufficiently large elliptic flow   to be accessible experimentally  in 4-th order cumulant analysis.

\end{abstract}


\date{\today}  

\maketitle

Recent experiments  at RHIC produces convincing evidences that a collective matter is created  
in Au+Au/Cu+Cu collisions \cite{BRAHMSwhitepaper,PHOBOSwhitepaper,PHENIXwhitepaper,STARwhitepaper}. The evidences come mainly from observing finite, large elliptic flow in
non-central collisions. Elliptic flow is the 2nd harmonic in the Fourier expansion of the momentum distribution of the identified particles, 

\begin{equation} \label{eq1}
\frac{1}{2\pi}\frac{dN}{p_Tdp_Tdyd\phi}=\frac{1}{2\pi}\frac{dN}{p_Tdp_Tdy}[1+2\sum_n v_n cosn(\phi-\phi_R)]
\end{equation}

\noindent $\phi_R$ being the azimuthal angle of the reaction plane. Elliptic flow ($v_2$) measure the azimuthal correlation of produced particles with respect to the reaction plane. Finite elliptic flow is now regarded as a definitive signature of collective effect  \cite{Ollitrault:1992bk,Poskanzer:1998yz}.  It is also best understood in a collective model like hydrodynamics \cite{QGP3}. 
In a non-central collision, the reaction zone is spatially asymmetric. Differential pressure gradient convert the spatial asymmetry in to momentum asymmetry. In other words, in a hydrodynamic model, spatial asymmetry of the interaction region produces collective effects.
Now   protons also have finite extension (though of smaller size) as do a nucleus. It is then possible that  in finite impact parameter pp collisions at the Large Hadron Collider (LHC), at c.m. energy $\sqrt{s}$=14 TeV, asymmetric reaction zone will produce  collective behavior which could be manifested as elliptic flow. 
However, even if flow is produced, whether or not it will be accessible experimentally will depend 
 on both the flow strength and the multiplicity in the phase space window where the flow is measured.
This is because non-flow effects like di-jet production, also show azimuthal correlation not related to the reaction plane. They need to be disentangled for faithful reconstruction of the reaction plane. Several standard methods  \cite{Poskanzer:1998yz,Ollitrault:1993ba,Borghini:2001vi,Bhalerao:2003xf} have been devised to discriminate non-flow effects. Event plane method \cite{Poskanzer:1998yz,Ollitrault:1993ba} determine the reaction plane, but require large multiplicity for unambiguous determination. 
Cumulant method \cite{Borghini:2001vi} does not require measurement of the reaction plane. Cumulants of multiparticle azimuthal correlation are related to flow harmonics.  The cumulants can be constructed in increasing order according to the number of particles that are azimuthally correlated. The method relies on the different 
multiplicity scaling property of the azimuthal correlation related to flow and non-flow effects. In the cumulant method, for charged particle multiplicity $n_{mult}$, $v_2$ can be reliably extracted using two particle correlator, if $v_2\{2\} >1/\sqrt{n_{mult}}$. Higher order correlators will increase the sensitivity, e.g.
$v_2\{4\} > 1/n_{mult}^{3/4}$. Maximum sensitivity $v_2 >1/n_{mult}$ could be achieved using still higher order correlators (cumulants of order greater than 4). 
%Added in the revised version
In the Lee-Yang zero method \cite{Bhalerao:2003xf} elliptic flow is obtained from
the zeros in a complex plane of a generating function of azimuthal correlation.
It is also less biased by the non-flow correction, $v_{2}>1/n_{mult}$. It may be mentioned here that the sensitivity arguments are based on order of magnitude and are mostly valid in large multiplicity limit.

Simulations of pp collisions at LHC 
energy by event generators like PYTHIA, indicate that while charged particle multiplicity in central rapidity region peaks at $n_{mult} <$ 10, the distribution has a pronounced high multiplicity tail. There are 
appreciable number of events with multiplicity $n_{mult} >50$, comparable to multiplicity in
peripheral Au+Au  collisions at RHIC. 
If  high multiplicity ($n_{mult}>50$) pp collisions generate elliptic flow $v_2 > 0.15$, flow would
be experimentally accessible in the 2nd or higher order cumulant analysis.
However, bulk of the pp collisions produces multiplicity $n_{mult} \sim$ 10. For $n_{mult}\sim10$, one readily gets $v_2\{2\} >0.3$, $v_2\{4\} > 0.17$, $v_2\{>4\} >0.1$ as the scales of elliptic flow needed to be produced 
to be experimentally accessible with 2nd, 4th and higher  order cumulant method.
Note that
the minimum $v_2$, $v_2\sim 0.1$ that can be accessed experimentally (with higher order cumulants)  is rather large. Peripheral Au+Au collisions at RHIC energy produces much less flow, $v_2\sim 0.06$ \cite{Back:2004mh}.
 
 In the present paper, in a hydrodynamic model, we explore the possibility of observing rather large elliptic flow ($v_2 \geq$ 0.1) in low multiplicity, $n_{mult}\sim$ 10, pp collisions at LHC. Applicability of hydrodynamics in a small   system like pp is uncertain.
Hydrodynamics require 'local' thermal equilibration, which can be achieved only if the mean free path of the constituents is small compared to the size of the system, $\lambda << R$.   In pp collisions, size of the system is not large, $\lambda \sim R \sim$ 1 fm. However, it can be argued \cite{Arnold:2004ti} that the essential assumption of ideal hydrodynamics is  
that the stress tensor is isotropic in the local rest frame, $T_{ij}=p\delta_{ij}$, with some equation of state relating pressure $p$ to energy density. The condition $T_{ij}=p\delta_{ij}$ is a statement of isotropisation of the medium. If the medium is isotropised within a time scale $\tau_i$, hydrodynamic may be applicable 
beyond $\tau_i$.
 In a QGP, non-abelian version of   Weibel instabilities can grow very fast, isotropising the medium \cite{Arnold:2004ti}. The maximum growth rate is, $\gamma \sim g\sqrt{n/p_{hard}}$, $n$ being the density and $p_{hard}$ is the characteristic momentum scale dominating the  excitations in non-equilibrium QGP
 \cite{Arnold:2004ti}. In the saturation scenario, $p_{hard}  \sim Q_s$, the saturation scale \cite{Baier:2000sb}. If one further assume, $n\sim Q_s^3/g^2$, then $\gamma \sim Q_s$, so the typical isotropization time scale is $1/Q_s$.
 In a central (b=0) collision, this scale in proton is larger than in a nucleus,
 $Q_{s,A} \geq  Q_{s,p}$.   \cite{Kowalski:2003hm}. Taking $Q_s\sim 1 GeV$,  one obtains the isotropization time scale as $\sim$ 0.2 fm, justifying appllicability of hydrodynamics in central pp collisions after $\tau_i\approx 0.2 fm$.
Indeed, similarities between pp and Au+Au collisions has been observed even at RHIC energy \cite{Chajecki:2009es}.
When phase space restriction due to conservation laws is taken into account
transverse momentum distribution in pp and Au+Au collisions at RHIC energy  show similar behavior \cite{Chajecki:2009es}. However, similarity in $p_T$ spectra alone does not prove that collective model like hydrodynamic is applicable in pp collisions. Observation of finite elliptic flow could be a definitive signature of collective behavior in pp collisions. Recently, several authors  \cite{Prasad:2009bx,CasalderreySolana:2009uk,Bozek:2009dt}  have applied ideal hydrodynamics to 
study elliptic flow in pp collisions at LHC. In \cite{CasalderreySolana:2009uk,Bozek:2009dt}, it was conjectured that high multiplicity events will show large elliptic flow, $v_2 \sim$ 10-20\% if hot spot like structures are produced in the initial collisions.
In \cite{Prasad:2009bx}, smooth initial energy density configuration was evolved hydrodynamically. For energy density appropriate for pp collisions at LHC, in peripheral collision small elliptic flow $v_2 \sim$ 2-3\% , was predicted.
With smooth energy density, hydrodynamical evolution donot generate elliptic flow in central collisions. 
 
Presently, we assume that in a 'central' pp collision at LHC energy, a number $N_s$ of 'hot spots' are created. Recently, in \cite{CasalderreySolana:2009uk,Bozek:2009dt}, scenarios with hot spots formation are considered. 
We do not dwell on the mechanism of hot spots formation. In the constituent quark model, only a few partons (quarks) interact,
forming  flux tubes between them. The hot spots can be formed from the rapid decay of the flux tubes \cite{Bozek:2009dt}. However, relevance of constituent quarks in LHC energy collisions is questionable. The collisions dynamics will be governed mainly by gluons. Large (gluon) density fluctuations can  lead to formation of hot spots \cite{CasalderreySolana:2009uk}. Hot spots are assumed to have Gaussian density distribution with $\sigma$=0.3 fm. For $N_s$ number of hot spots, the energy density of the system can be obtained as,
\begin{figure}[t]
\vspace{0.35cm} 
\center
 \resizebox{0.3\textwidth}{!}{%
  \includegraphics{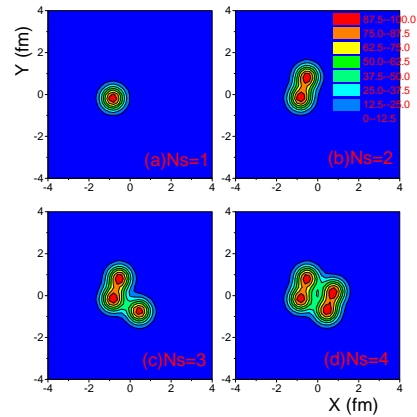} 
}
%\vspace{-4cm}
\caption{  (color online) Transverse profile of the energy density in a random event, with (a) one, (b) two,
(c) three and (d) four   hot spots, are shown. Peak energy density is assumed to be $\varepsilon_0$=100 $GeV/fm^3$.   }\label{F1}
\end{figure}

\begin{equation}
\varepsilon({x,y})=\varepsilon_0 \frac{1}{\sqrt{2 \pi \sigma^2}}
\sum_{i=1}^{i=N_s} e^{-\frac{({\bf r}-{\bf r}_i)^2}{2\sigma^2}}
\end{equation}

The centre of the hot spots (${\bf r}_i$) can be anywhere in the reaction volume. Hydrodynamic evolution of the fluid will depend on the positions (${\bf r}_i$). In the following, we assume that (${\bf r}_i$)'s are 
randomly distributed within a  sphere of radius R=1.2 fm. In Fig. \ref{F1}, 
  transverse energy density profile of a random event, with one, two, three and four      hot spots are shown in four panels.
The peak energy density  is assumed to be $\varepsilon_0$= 100 $GeV/fm^3$. As it will be discussed below, for $\varepsilon_0$=100 $GeV/fm^3$, boost-invariant hydrodynamical evolution of   2-3  hot spots, produces charged particle multiplicity in the range  $n_{mult}=8-10$, close to the number obtained in simulations with event generators.

 From Fig.\ref{F1}, it is obvious that depending on the number of hot spots, initial spatial asymmetry or the participant eccentricity will  vary. For a finite number of hot spots,  participant eccentricity can be obtained  as  \cite{Alver:2008zza},
 
\begin{equation}
\epsilon=\frac{\sqrt{(\sigma_y^2-\sigma_x^2)^2-4\sigma_{xy}}}{\sigma_y^2+\sigma_x^2} 
 \end{equation}
 
\noindent where $\sigma_x^2=\la x^2\ra-\la x\ra^2$, $\sigma_x^2=\la x^2\ra-\la x\ra^2$, $\sigma_y^2=\la y^2\ra-\la y\ra^2$, $\sigma_{xy}=\la xy\ra-\la x\ra \la y\ra$ and $\la...\ra$ denote density weighted averaging. In the limit of homogeneous density distribution, $\epsilon$ coincides with the usual 
 definition of participant eccentricity \cite{Poskanzer:1998yz}, $\epsilon\prime=\frac{\la y^2\ra-\la x^2\ra}{\la y^2\ra+\la x^2\ra}$. In table.\ref{table1},
  we have noted the participant eccentricity as a function of the number of hot spots. They are averaged over 200 random initial configurations. Error are statistical only. When only a single hot spot is formed, $\epsilon$ is consistent with zero. Participant eccentricity is finite if 2 or more hot spots are formed. It also appear that for more than one hot spots, participant eccentricity is approximately constant, $\epsilon \sim 0.5$, which is comparable to that in very peripheral Au+Au collisions.

\begin{figure}[t]
\vspace{0.3cm} 
\center
 \resizebox{0.35\textwidth}{!}{%
  \includegraphics{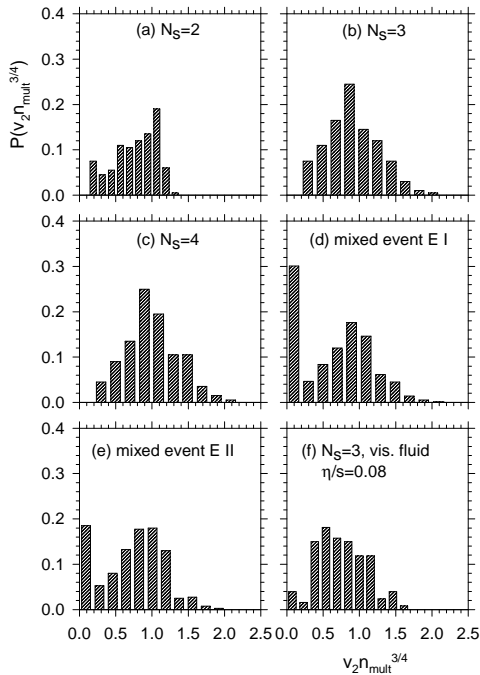}%{pdistr.eps}  
}
%\vspace{-4cm}
\caption{Probability distribution $P(v_2 n^{3/4}_{mult})$ in simulations with two, three and four hot spots are 
are shown in panels (a-c). $P(v_2 n^{3/4}_{mult})$ in mixed events I and II are shown in panels (d) and (e). In panel (f),  $P(v_2 n^{3/4}_{mult})$ in 
 viscous ($\eta/s$=0.08) fluid evolution, with three hot spots, is shown. 
  }   \label{F2}
\end{figure}

We assume that 'baryon free'  hot spots are 'locally' thermalised in the time scale $\tau_i$=0.2 fm and evolve hydrodynamically.  
The space time evolution of the fluid is then obtained by solving the energy-momentum conservation equation,

\begin{equation}  \label{eq2}
\partial_\mu T^{\mu\nu} =  0,\\  
\end{equation}

\noindent where $T^{\mu\nu}=(\varepsilon+p)u^\mu u^\nu - pg^{\mu\nu}$, is the energy-momentum tensor.
$\varepsilon$, $p$ and $u$ being the energy density, pressure and fluid velocity respectively. 
Assuming boost-invariance, Eqs.\ref{eq2}  is solved in $(\tau=\sqrt{t^2-z^2},x,y,\eta_s=\frac{1}{2}\ln\frac{t+z}{t-z})$ coordinates, with a code 
  "`AZHYDRO-KOLKATA"', developed at the Cyclotron Centre, Kolkata. Initial   fluid velocity was assumed to be zero. 
 Details of the code can be found in  \cite{Chaudhuri:2008sj,Chaudhuri:2008ed,Chaudhuri:2009uk}.  
Solution of Eq.\ref{eq2} requires an equation of state (EoS). We assume a lattice based EoS   with confinement-deconfinement cross over transition at $T_{co}$=196 MeV.
Details of the EoS can be found in    \cite{Chaudhuri:2009uk,Chaudhuri:2009ud}.  
  Recent lattice simulations \cite{Cheng:2007jq} have been parameterised to obtain EoS for the deconfined phase.
 EoS of the confined phase is that of a  
non-interacting hadronic resonance gas comprising resonances with mass $m < 2.5 $ GeV. 
Hydrodynamic evolution of initial QGP fluid, with the lattice based EoS correctly reproduces large volume of experimental data in Au+Au collisions at RHIC \cite{Chaudhuri:2009uk,Chaudhuri:2009ud}.

  \begin{table}[t] 
\caption{\label{table1} Event averaged (for sample size of 200) participant eccentricity ($\epsilon$) are shown as a function of number of  hot spots
($N_s$). Also shown are the event averaged charged particle multiplicity ($\la n_{mult}\ra)$, mean transverse momentum ($\la p_T\ra$) and elliptic flow ($v_2$)   as a function of number of hot spots ($N_s$). The bracketed quantities are the averages when sample size is halved. Last two rows corresponds to mixed events EI and mixed events EII (see text).}
\begin{tabular}{|c|c|c|c|c|}\hline
  $N_s$   & $\epsilon$ &$\la n_{mult}\ra$ & $\la p_T\ra (GeV)$  & $\la v_2\ra$ \\  
  \hline
1 &0 & $4.97\pm 0.02$ & $0.722\pm 0.001$ & $0.003\pm 0.001$   \\  %\hline
  &  & $4.97\pm 0.02$)&($0.722\pm 0.001$)&($0.003\pm 0.001$)  \\  \hline
2 &$0.532\pm 0.052$ & $7.75\pm 1.17$ & $0.634\pm 0.054$ & $0.147\pm 0.071$ \\   
  &  &($7.88\pm 1.11$)&($0.632\pm 0.054)$& $0.152\pm 0.068$) \\  \hline
3 & $0.536\pm 0.051$ & $9.68\pm 2.24$ & $0.599\pm 0.037$ & $0.160\pm 0.053$     \\  %\hline
 &   & ($9.87\pm 2.12$) & ($0.601\pm 0.040$) & ($0.158\pm 0.056$)    \\  \hline
4 & $0.457\pm 0.048$ & $11.05\pm 2.58$  & $0.582\pm 0.029$ & $0.161\pm 0.050$   \\  %\hline
  &                  & ($11.39\pm 2.67$)  & ($0.581\pm 0.026$) & ($0.160\pm 0.049)$   \\  \hline
EI &   & $8.36\pm 2.91$  & $0.634\pm 0.065$ & $0.118\pm 0.019$   \\  \hline
EII &   & $8.45\pm 2.36$  & $0.627\pm 0.057$ & $0.138\pm 0.022$   \\  \hline
\end{tabular} 
\end{table} 
  
For a given hot spots configuration, hydrodynamic equations are solved to obtain freeze-out surface at a fixed temperature $T_F$=130 MeV. Using the Cooper-Frey prescription,
invariant distribution of $\pi^-$ is obtained. In the present paper, we have neglected resonance contribution. We multiply the $\pi^-$
multiplicity by a factor of 1.5 to approximately account for the resonance contribution and noting that only $\sim$80\% of charged particles are pions, it is further multiplied by a factor of $2\times 1.2$ to obtain the charged particle multiplicity. In table.\ref{table1},   event averaged charged particle multiplicity $\la n_{mult}\ra$,
mean transverse momentum  $\la p_T \ra$, and transverse momentum integrated elliptic flow $\la v_2\ra$ 
are noted as a function of the number of hot spots. Sample size is $N_{event}$=200. The errors shown are statistical. In table.\ref{table1}, we have also noted the averages when the event size is halved (the bracketed values). 
It is worth noting that even if the sample size is halved, averages and statistical error remain approximately unchanged. This indicates that the   sample size is statistically large and simulation results are stable.
The general trend of the $\la n_{mult}\ra$, $\la p_T\ra$, and $\la v_2\ra$   are well established.
Multiplicity increases with number of hot spots $N_s$, e.g. as $N_s$ increases from 1 to 4, $\la n_{mult}\ra$ increases by a factor of $\sim$ 2. 
   The result is understood. We have fixed the peak energy density $\varepsilon_0$=100 $GeV/fm^3$.
As the number of hot spots increases, initial total energy of the system also increases, and so does the multiplicity. For $N_s$=1, the average multiplicity 
$\la n_{mult}\ra \sim 4$. Average multiplicity is nearly a factor of 2 less than the peak multiplicity ($\sim$7-8) expected in pp collisions. However,  multiplicity will increase if $\varepsilon_0$ is increased and   expected charged particles multiplicity
could be reproduced even with a single hot spot. In pp collisions, one expects 2-3 hot spots are formed. 
For 2-3 hot spots with peak energy density  100 $GeV/fm^3$, average number of charged particle multiplicity $\la n_{mult}\ra=8-10$ is close to the expected multiplicity.  While $\la n_{mult}\ra$ increases with number of hot spots, 
mean transverse momentum of charged particles decreases. 
  However, for two or more hot spots, the decrease is less than 10\%. It appears that the slope of the transverse distribution of charged particles will remain approximately constant if two or more   hot spots are formed. Simulation results for elliptic flow are most interesting. Determination of
elliptic flow requires the azimuthal orientation of the reaction plane. For a number of hot spots, arbitrary  located, the azimuthal orientation of the reaction plane is non-trivial. We calculate  $v_2$   assuming that the reaction plane is oriented such that the $v_2$ is maximised. As expected, $\la v_2\ra$ is consistent with zero  if a single hot spot is produced in the collisions. For a single hot spot,  participant eccentricity is approximately zero and hydrodynamic evolution does not generate any flow.  
%%%%%%% added in the revised version %%%%%%%%%%
Indeed, events with a single hot spot   essentially corresponds to smooth hydrodynamics, and with respect to elliptic flow development, are equivalent to events without any hot spot. If single hot spot formation or smoothed hydrodynamics dominates,  elliptic flow will not be observed in central pp collisions.
%%%%%%%%%%%%%%%%%%%%%%%%%%
If in pp collisions  more than one hot spot is formed, then substantial flow is generated.  
 It also appear that, for two or more hot spots, $\la v_2\ra$ is approximately constant, $\la v_2\ra \sim 0.15 \pm 0.05$. With two or more hot spots in the initial state, central pp collisions at LHC could generate more elliptic flow  than that generated in peripheral Au+Au collisions at RHIC.    
 
We have also considered mixed events, (i) EI, when probability of formation of one, two, three and four hot spots are equal and (ii) EII, when probability of formation of one, two, three and four hot spots are 10\%, 50\%, 30\% and 10\% respectively. In a realistic situation mixed events EII are more likely than EI. 
In table.\ref{table1}, in last two rows, $\la n_{mult}\ra$, $\la p_T\ra$, and $\la v_2\ra$ for the mixed events EI and EII are noted. In mixed events EI and EII, elliptic flow is still large, $\la v_2 \ra \sim$ 10-15\%.

Simulation results indicate that if more than one hot spot is formed in initial
pp collisions, large elliptic flow may result in low multiplicity events.
However, generating large elliptic flow  does not ensure that they will be accessible in experiment. As discussed earlier, whether or not the flow is accessible experimentally depend on both the flow strength ($v_2$) and multiplicity ($n_{mult}$). For $n_{mult}\sim$10, event plane method \cite{Poskanzer:1998yz,Ollitrault:1993ba} of determination of the reaction plane will have very large uncertainty. Cumulant method \cite{Borghini:2001vi} do not measure the reaction plane.  
To find the order of cumulant required for faithful measurement of $v_2$,  we computed the measure (i)   $v_2.\sqrt{n_{mult}}$ and (ii) $v_2.n^{3/4}_{mult}$.
Simulated events donot satisfy the condition $v_2.\sqrt{n_{mult}}>1$, necessary for two particle correlator to measure $v_2$. 
In Fig.\ref{F2} probability distribution $P(v_2.n^{3/4}_{mult})$ of $v_2.n^{3/4}_{mult}$ in the simulated events are shown.
In panels (a-c) distribution obtained with two, three and four hot spots are shown. $P(v_2.n^{3/4}_{mult})$ in mixed events I and II are shown in panel (d) and (e).  
In all the cases, we find that more than 20\% events crosses the threshold,
$v_2.n^{3/4}_{mult}> 1$ for 4th order cumulant analysis.

In the present simulations, we have neglected the effect of viscosity. 
Recently, from a systematic analysis of STAR data on $\phi$ mesons, QGP viscosity was estimated, $\eta/s=0.07 \pm 0.03 \pm 0.14$ \cite{Chaudhuri:2009uk}, the first error is statistical and the second one is systematic. The central value of the estimate is close to the ADS/CFT lower bound on viscosity, $\eta/s \geq 1/4\pi$  \cite{Policastro:2001yc}. 
  Other parameters remaining unchanged, effect of viscosity is to reduce elliptic flow and increase particle multiplicity. For example,  
in Au+Au collisions, compared to ideal fluid evolution, in viscous fluid ($\eta/s=1/4\pi$) evolution,   elliptic flow is reduced by $\sim$10\%, and particle multiplicity is increased by $\sim$ 20\%  \cite{Chaudhuri:2009uk}. However, if hot spots are formed in pp collisions, the gradients will be steeper and     it is possible that viscous effects will be  more than that in   nuclear collisions. Indeed, explicit simulations also indicate that viscous effects are more in pp collisions.
For example, in a typical pp event, with three hot spots, with peak energy density, $\varepsilon_0$=100 $GeV/fm^2$, compared to ideal fluid evolution, in minimally viscous ($\eta/s$=0.08) fluid evolution, elliptic flow is reduced by $\sim$30\%, and particle multiplicity is increased by $\sim$80\%. If one adjust the initial energy density to obtain multiplicity ($n_{mult}\sim$10), as in ideal fluid evolution, then   elliptic flow is reduced by $\sim$15\%.  
In Fig.\ref{F2}f, we have shown the probability distribution, $P(v_2.n^{3/4}_{mult})$, in minimally  viscous ($\eta/s$=0.08) fluid evolution when three shot spots are formed in initial pp collisions. The peak energy density is adjusted to $\varepsilon_0$=50 $GeV/fm^3$, such that average particle multiplicity is $n_{mult}=11.39 \pm 0.83$.   Approximately 15\% of the events satisfy the criterion, $v_2.n^{3/4}_{mult}>1$.

%%%%%%%%%%%%%Added in the revised version %%%%%%%%%%%%%%%%%%% 
In the present simulations, we have considered only hot spots with size $\sigma$=0.3 fm. 
The size of the hot spots is important for the development of elliptic flow,
and it is important to understand the effect on simulations  
if hot spot sizes are changed. In a hydrodynamic model, elliptic flow depends on the initial spatial eccentricity. Initial spatial eccentricity increases if hot spot size is reduced \cite{CasalderreySolana:2009uk}. Elliptic flow in the model will increase if hot spot size is reduced. Explicit simulation indicate that elliptic flow will increase by $\sim$10\% if hot spot size is reduced from 0.3 to 0.2 fm. The criterion $v_2n_{mult}^{3/4} > 1$ will be better observed then. On the contrary, if hot spot size is large, $v_2$ decreases and the criterion $v_2n_{mult}^{3/4} > 1$    will be less well observed.
%%%%%%%%%%%%%%%%%%%%%%%%%%%%%%%%%%%%%%%%%%%%%%%%%%%%%%%%%%%%%%%

%%added in the re-revised version.
Before we summarise our results, we would like to comment on  'typicality' of hot spot formation. We have considered central pp collisions. Central collisions do not dominate the cross section.  If events with at least two hot spot are only a small sub set of already rare central events, then the present result, large elliptic flow in central pp collisions due to hot spot formation, may not be relevant experimentally. We argue here that two hot spot formation probability is not small, rather it is large. Hot spot formation is possibly connected with multiple interactions of partons. For example, double partonic scattering (DPS) may lead to 
formation of two hot spots. In hadronic collisions,  evidences of double parton scattering have been found in multi jet events  \cite{Alitti:1991rd,Abe:1993rv}. Double parton scatering events are not rare. For example, 
CDF collaboration \cite{Abe:1993rv} estimated the double parton scattering events as  $\sim$ 52\% in $p+\bar{p}\rightarrow \gamma + 3 \text{jets} +X$ reactions at $\sqrt{s}$=1.8 TeV. Theoretical investigations also predict large number of double parton events 
at LHC energy    \cite{Sjostrand:2004pf,Corke:2009tk}. In minimum bias collisions, more that 50\% double parton scattering are predicted. In events with hard process, the fraction is even larger.

To summarise, due to finite extension of protons it is possible that at LHC energy ($\sqrt{s}$=14 TeV),  pp collisions will produce  collective matter. Recently in \cite{CasalderreySolana:2009uk,Bozek:2009dt}, it was %has been
conjectured that if pp collisions led to formation of hot spot like structures, in high multiplicity  $n_{mult} > 50$ events,   collective effects will be manifested as 'experimentally measurable' elliptic flow. Bulk of the pp collisions are low multiplicity ($n_{mult}\sim 10$) events. 
We have explored the possibility of observing elliptic flow in low multiplicity events. In a hydrodynamic model, we have simulated pp collisions. It is assumed that initial collisions led to formation of a number of hot spots, distributed randomly in the interaction volume. Hydrodynamical evolution of   two or more hot spots with peak energy density $\varepsilon_0$=100 $GeV/fm^3$,  produces  charged particle multiplicity, $n_{mult}\approx 8-10$ as expected in bulk of the events in pp collisions.
Formation of   
  two or more hot spots also generate substantial elliptic flow,
  $v_2\approx 0.15 \pm 0.05$. Large elliptic flow in low multiplicity pp collisions will be accessible experimentally in  4th order cumulant analysis.

\end{document}